\documentclass[aps,twocolumn,amssymb,amsmath]{revtex4}
\usepackage{graphicx}
\begin{document}

\title{Field test of the wavelength-saving quantum key distribution network}

\author{Shuang Wang,$^{1}$ Wei Chen,$^{1,*}$ Zhen-Qiang Yin,$^{1}$ Yang
Zhang,$^{1}$ Tao Zhang,$^{2}$ Hong-Wei Li,$^{1}$ Fang-Xing Xu,$^{1}$
Zheng Zhou,$^{1}$ Yang Yang,$^{2}$ Da-Jun Huang,$^{2}$ Li-Jun
Zhang,$^{2}$ Fang-Yi Li,$^{1}$ Dong Liu,$^{1}$ Yong-Gang Wang,$^{2}$
Guang-Can Guo,$^{1}$ and Zheng-Fu Han$^{1,*}$}

\address{
$^1$Key Laboratory of Quantum Information, University of
Science and Technology of China, CAS, Hefei 230026, China \\
$^2$Laboratory of Fast Electronics, University of Science and
Technology of China, Hefei 230026, China\\
$^*$Corresponding authors: kooky@mail.ustc.edu.cn, zfhan@ustc.edu.cn
}

\begin{abstract}
We propose a wavelength-saving topology of quantum key distribution
(QKD) network based on passive optical elements, and report the
field test of this network on the commercial telecom optical fiber.
In this network, 5 nodes are supported with 2 wavelengths, and every
two nodes can share secure keys directly at the same time. All QKD
links in the network operate at the frequency of 20 MHz. We also
characterized the insertion loss and crosstalk effects on the
point-to-point QKD system after introducing this QKD network.
\end{abstract}

\maketitle

Quantum key distribution enables separate legitimate participants to
share keys with provable unconditional security \cite{gisin02}.
After last two decades of developments, the point-to-point (P2P)
QKD, which only makes two separate parties to share secure keys, has
matured sufficiently to provide a practical secure communication,
some companies have developed their related commercial products.
Furthermore, the QKD network was proposed to satisfy future demands
for large-scale and multi-user secure communication. Since Townsend
et al. presented and realized the first QKD network
\cite{townsend94,townsend97}, various QKD network topologies have
been designed and subsequently demonstrated
\cite{BBN,WDM,bus,zhangtao,chenwei,SECOQC08,pan,xu,SECOQC09}, in
which refs \cite{BBN,chenwei,SECOQC08,pan,xu,SECOQC09} demonstrated
the QKD network on real-built telecom optical fiber networks.

Based on the wavelength division multiplexing (WDM), we have
designed a real-time-full-connectivity (RTFC) QKD network, in which
every two nodes in the network can share secure keys directly at the
same time \cite{zhangtao}. In this paper we propose and field test a
new wavelength-saving RTFC QKD network. The QKD network with new
topology can support $2N+1$ nodes only with $N$ wavelengths, which
saves about $50\%$ of the wavelengths compared with the previous
topology.

\begin{figure}
\begin{center}
\includegraphics[width=0.35\textwidth]{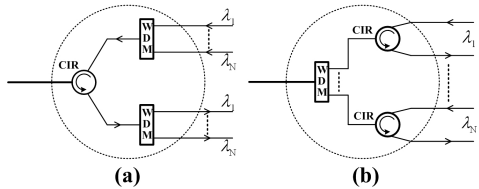}
\caption{Two structures of the basic unit with 3-port CIRs and
$N$-wavelength WDMs. $\lambda_{1}\cdots\lambda_{N}$ are the $N$
wavelengths band of the WDM and arrows mark directions.}
\label{graph1}
\end{center}
\end{figure}
The wavelength-saving topology employs the basic unit consisting of
one 3-port circulator (CIR) and two $N$-wavelength WDMs
(Fig.\ref{graph1}(a)), or one $N$-wavelength WDM and $N$ 3-port CIRs
(Fig.\ref{graph1}(b)). This basic unit can be regarded as a
multiplexer joining $N$-wavelength input signals together and a
demultiplexer splitting $N$-wavelength output signals apart, denoted
as M\&D.
\begin{figure}
\begin{center}
\includegraphics[width=0.35\textwidth]{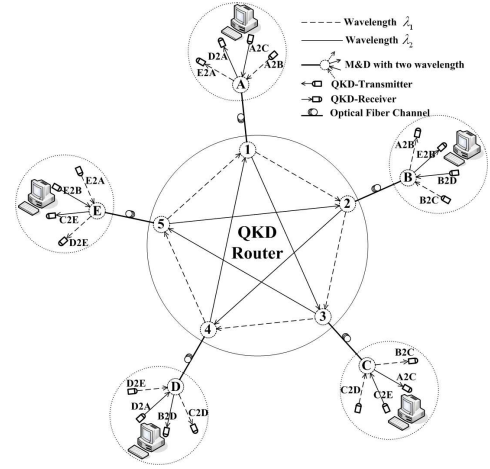}
\caption{Schematic diagram of the 5-node QKD network topology with 2
wavelengths. \textbf{A}, \textbf{B}, \textbf{C}, \textbf{D}, and
\textbf{E} are the five nodes, and \textbf{1}, \textbf{2},
\textbf{3},\textbf{4}, and \textbf{5} are the five ports of the QKD
router. The arrows indicate the propagation direction of photons.}
\label{graph2}
\end{center}
\end{figure}

As an example, figure \ref{graph2} shows the topology of a 5-node
RTFC QKD network with 2 wavelengths. Its core component -- the
5-port QKD router is composed of five $1\times4$ M\&Ds. All nodes
connect to the router with $1\times4$ M\&Ds. When node \textbf{A}
wants to share secret keys with \textbf{B}, its QKD-Transmitter
\textbf{A}2\textbf{B} sends photons of wavelength $\lambda_1$ to its
M\&D which multiplexes these photons to the optical fiber channel,
while arriving at the router, these photons will propagate from port
\textbf{1} to \textbf{2}, then forward to node \textbf{B} and be
demultiplexed by the M\&D, and finally received by \textbf{B}'s
corresponding QKD-Receiver \textbf{A}2\textbf{B}. At the same time,
if node \textbf{A} want to share other keys with \textbf{D}, who
would be required to send photons of wavelength $\lambda_2$ to
\textbf{A}. Therefore \textbf{A} can transmit photons of wavelength
$\lambda_1$ and $\lambda_2$ to \textbf{B} and \textbf{C}, and
receive photons of wavelength $\lambda_2$ and $\lambda_1$ from
\textbf{D} and \textbf{E}, respectively and simultaneously. Every
node in this architecture is on the same term, so every two nodes
can share secure keys directly at the same time only with two
wavelengths. Since photons in the network propagate unidirectionally
(only from one node to the other), any unidirectional P2P QKD system
\cite{uni1,uni2,uni3,uni4,uni5,uni6} can be applied on this QKD
network independently. The QKD router can be extended to $2N+1$
ports with $N$ wavelengths based on the Hamiltonian circuits theorem
in odd complete graph theory (Theorem 2-8 in ref.\cite{graph}), then
the above example can be easily extended to a $2N+1$ nodes network.

\begin{figure}
\begin{center}
\includegraphics[width=0.35\textwidth]{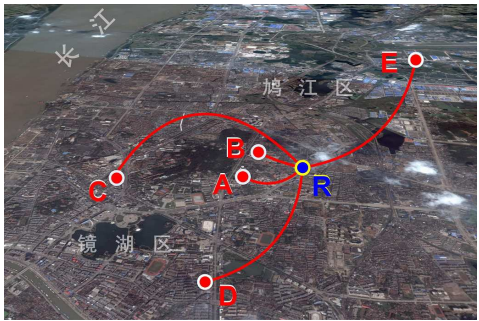}
\caption{Satellite map with the locations of five nodes in the QKD
network in Wuhu, Anhui, China. \textbf{R} stands for the QKD router,
which is settled in the Telecom Station.}\label{map}
\end{center}
\end{figure}

The wavelength-saving 5-node RTFC QKD network with 2 wavelengths
(Fig.\ref{graph2}) was field tested on the commercial fiber network
of China Telecom Corporation Ltd. in Wuhu, Anhui, China.
Fig.\ref{map} shows locations of 5 nodes and the QKD router in the
satellite map, and 25.28km Corning SMF-28e fiber was added behind
the commercial fiber at \textbf{E}. The routing rule was as follows:
\textbf{A}2\textbf{B}, \textbf{B}2\textbf{C},
\textbf{C}2\textbf{D},\textbf{D}2\textbf{E}, \textbf{E}2\textbf{A}
with $\lambda_1=1530nm$, and \textbf{A}2\textbf{C},
\textbf{C}2\textbf{E}, \textbf{E}2\textbf{B}, \textbf{B}2\textbf{D},
\textbf{D}2\textbf{A} with $\lambda_2=1550nm$.

Employing BB84\cite{bb84} phase coding protocol with 3-intensity
decoy state method\cite{ma}, the asymmetric Faraday-Michelson
interferometer (AFMI) setup\cite{uni5} was adopted to realize all
the P2P QKDs in the network. In each P2P QKD, photon pulses with
750ps duration are generated by a 1530nm (or 1550nm) pulsed laser
operating at 20 MHz, and modulated by a fiber-optic intensity
modulator (IM) to create signal, decoy and near vacuum sate, then
randomly phase-coded ($\{0, \pi\}, \{\frac{\pi}{2},
\frac{3\pi}{2}\}$) and phase-decoded ($\{0, \pi\}, \{\frac{\pi}{2},
\frac{3\pi}{2}\}$) by the AFMI pairs belonged to the QKD-transmitter
and QKD-receiver respectively, finally detected by single SPD. The
InGaAs SPDs produced by Princeton Lightwave have quantum efficiency
of about $20\%$ and dark counts of nearly $2\times10^{-5}$ per gate
at their maximum trigger frequency 20 MHz and 1ns gate duration. The
average photon number of signal state and decoy state are $\mu=0.6$
and $\nu=0.2$, and the ratio among the signal state, decoy sate and
vacuum state is $6:3:1$. Although the extinction ratio of IM is 27
dB, the maximum calculated yield of vacuum states
$Y_0=1.24\times10^{-5}$ among 10 links is below dark counts of SPDs,
so the near vacuum states are vacuum for SPDs.

\begin{table}[hbt]
\begin{center}
\caption[fake]{Field test results for the QKD network}
\label{result}
\begin{tabular}{c||c|c|c|c}
\hline \hline           &\textbf{A}2\textbf{R}2\textbf{B} &\textbf{A}2\textbf{R}2\textbf{C}  &\textbf{D}2\textbf{R}2\textbf{A}  &\textbf{E}2\textbf{R}2\textbf{A}  \\
\hline Wavelength (nm)  &1530                             &1550                              &1550                              &1530              \\
Attenuation (dB)        &7.24                             &8.78                              &10.79                             &14.77              \\
Crosstalk (dB)          &$-$38.37                         &$-$36.07                          &$-$35.88                          &$-$34.62              \\
Dead-time ($\mu$s)      &5                                &10                                &25                                &50                  \\
Sifted-key (kbit/s)     &31.00                            &17.64                             &8.16                              &3.83                \\
 Signal QBER (\%)       &2.92                             &2.84                              &2.78                              &3.76                \\
 Secure-key (kbit/s)    &4.91                             &2.02                              &1.82                              &0.41                \\
\hline\hline
\end{tabular}
\end{center}
\end{table}

Table \ref{result} shows parameters and results for QKD links
between \textbf{A} with other four nodes. The attenuation includes
the channel attenuation and effective insertion loss 3.10 dB (see
posterior paragraph) of network elements. Our control system provide
different dead-time for distinct QKD links to achieve optimal
secure-key rates. All the P2P QKD systems run over two hours, and
more than 115G pulses were sends out from each QKD-Transmitter. For
each P2P QKD link, the lower bound of the gain and the upper bound
of the error rate of single photon states were estimated according
to ref \cite{ma}, and then the secure key generation rate was
achieved by the GLLP formula \cite{gllp}. Here the statistic
fluctuation of measurement \cite{wangxb} were ignored. With the
measurement results in the QKD network, every two nodes could
ceaselessly share secure keys after post processing, then these
secure keys could be utilized for practical applications.

\begin{figure}
\begin{center}
\includegraphics[width=0.35\textwidth]{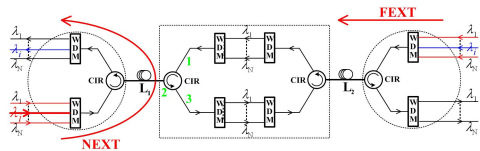}
\caption{Equivalent model of the QKD network topology. }
\label{graph4}
\end{center}
\end{figure}

In order to analyze effects on the P2P QKD system after introducing
the wavelength-saving QKD network, an equivalent model is proposed
(Fig.\ref{graph4}). The left and right M\&Ds stand for two nodes
respectively, the middle one composed by two M\&Ds is an equivalent
QKD router. For simplicity, four M\&Ds are assumed the same and each
one is composed of one CIR and two $N$-wavelength WDMs. In the
equivalent model, photons of wavelength $\lambda_i$ from the right
node to the left node (Blue in Fig.\ref{graph4}) are considered as
signal, while photons from other QKD-Transmitters (Red in
Fig.\ref{graph4}) are traded as crosstalk. Suppose the sifted-key
rate and quantum bit error rate (QBER) of the $\lambda_i$ signal are
$R_s$ and $QBER_0$ with the attenuation $\alpha_{_{Link}}$, the
insertion loss and crosstalk are two main performances we will
analyze.

 \emph{Insertion Loss} of QKD network elements and channel
attenuation compose the attenuation of each QKD link. Compared with
P2P QKD systems, four M\&Ds are added in the network. However, when
we calculate the insertion loss, only three M\&Ds need to be
considered because the multiplexer part of the M\&D following the
QKD-Transmitter can be regarded as part of the QKD-Transmitter
itself. After introducing the QKD network, the channel attenuation
reduced to $\alpha_{_{Link}}- \alpha_{_{IL}}$ to get the same $R_s$,
where $\alpha_{_{IL}}$ is the effective insertion loss. We measured
the insertion loss by connecting those five M\&Ds and the 5-port QKD
router with five 12m-long SMF patchcords, the effective insertion
loss is $\frac{3}{4} \times 4.14\approx3.10$ dB on average.

\emph{Crosstalk} originates from imperfections of network elements
and the channel. Since the SPD works in the Geiger mode, we
classified crosstalk into point one and continuous one, where the
point crosstalk comes from the return loss and directivity of CIRs
and back-reflection of joining interfaces between fibers, and the
continuous crosstalk is the Rayleigh backscattering in fibers. The
interband part, in which the signal and crosstalk have different
wavelengths, of point and continuous crosstalk is high order
crosstalk in this QKD network, and can be removed by
narrow-linewidth lasers and narrow-band filters. Therefore, only the
intraband part, in which the signal and crosstalk have the same
wavelength, need to be considered. And this part is totally
phase-independent of the signal because photons of crosstalk and
signal come from two remote lasers. So, the crosstalk considered
here is incoherent, and the increased QBER after introducing the QKD
network can be expressed as
\begin{equation}\label{crosstalk}
\Delta_+ QBER = \frac{\chi}{2}\cdot
\frac{1-QBER_0(3-2QBER_0)}{1+\chi (1-QBER_0)} < \frac{\chi}{2},
\end{equation}
where $\chi$ is the crosstalk ratio, and $\chi_{_{B}} \leq \chi \leq
\chi_{_{W}}$. We get the worst crosstalk $\chi_{_{W}}$ if all point
crosstalk fall in the SPD's gate duration. However, we can reduce
the crosstalk by adding delay in the node's $\lambda_i$
QKD-transmitter to adjust the arrival time of the point crosstalk,
and get the best one $\chi_{_{B}}$ if all point crosstalk fall out
of the SPD's gate duration.

The crosstalk in Table \ref{result} was measured as following: every
QKD-Transmitter of 5 nodes in Fig.\ref{graph2} was replaced by the
corresponding pulse laser, whose average power was adjusted to
$P_0=-24.00 dBm$, when the crosstalk of column
\textbf{A}2\textbf{R}2\textbf{B} was measured, we only shut down the
1530nm laser of node \textbf{A} and measured the output power
$P_{\textbf{A}2\textbf{R}2\textbf{B}}$ of 1530nm output port
belonged to node \textbf{B}, so the crosstalk was
$P_{\textbf{A}2\textbf{R}2\textbf{B}} - (P_0- \frac{1}{4}\times4.14)
$ (dB), the crosstalk of other columns could be measured in a
similar way. Then, the QKD link \textbf{E}2\textbf{R}2\textbf{A},
which has the largest attenuation and crosstalk, was chosen to
analyze effects of crosstalk in the field experiment. Considering
the worst case, in which all the point crosstalk fall into the gate
duration of the SPD, although $-34.62 > -27 - 14.77$, the average
gain of the crosstalk $Y_X=7.98\times10^{-6}$ is still below the
dark count of the SPD, and the $\Delta_+ QBER$ of the signal state
and the decoy state are $0.19\%$ and $0.25\%$ respectively, only
$\frac{1}{20}$ of either $QBER_0$. So, the effects of crosstalk in
our field experiment are negligible.

In summary, a wavelength-saving RTFC QKD network has been proposed
and field tested at 20 MHz. The experimental results validate that
the new topology is feasible for integration into the existing
telecom fiber network, and the insertion loss and crosstalk analysis
indicates that effects introduced by the QKD network are acceptable.

This work was supported by Wuhu Government and China Telecom
Corporation Ltd., Wuhu Branch, and the National Fundamental Research
Program of China (Grant No. 2006CB921900), the National Natural
Science Foundation of China (Grants No. 60537020 and No. 60621064),
the Innovation Funds of Chinese Academy of Sciences, and
International Partnership Project.

\makeatletter

\end{document}